\documentclass[useAMS,usenatbib,letterpaper]{mn2e}
\usepackage{graphicx,amsmath,color,amssymb,hyperref,fixltx2e}

\newcommand{\adsurl}[1]{\href{#1}{ADS}}

\newlength{\narrowfigurewidth}
\newlength{\figurewidth}
\newlength{\widefigurewidth}
\newcommand{\etal}
  {et al.}

\newcommand{\Lya}{Lyman-$\alpha\;$}
\newcommand{\Mpc}{\mathrm{Mpc}}
\newcommand{\Mpch}{\mathrm{Mpc} \,h^{-1}}

\title
  [$P(k)$ Reconstruction from Lyman-$\alpha$]
  {Minimally Parametric Power Spectrum Reconstruction from the Lyman-$\alpha$ Forest}

\author
  [S. Bird \etal]
  {Simeon Bird $^{1}$\thanks{E-mail: spb41@ast.cam.ac.uk},
  Hiranya V. Peiris $^{1,2}$,
  Matteo Viel $^{3,4}$,
  Licia Verde $^{5}$
\vspace{7mm}\\
$^1$Institute of Astronomy and Kavli Institute for Cosmology, Madingley Road, Cambridge CB3 0HA, U.K.\\
$^2$Department of Physics and Astronomy, University College London, London WC1E 6BT, U.K.\\
$^3$INAF - Osservatorio Astronomico di Trieste, Via G.B. Tiepolo 11, I-34131 Trieste, Italy \\
$^4$INFN/National Institute for Nuclear Physics, Via Valerio 2, I-34127 Trieste, Italy\\
$^5$ICREA \& Instituto de Ciencias del Cosmos, Universitat de Barcelona, Marti i Franques 1, 08028, Barcelona, Spain
}

\begin{document}

\date{}

\pagerange{\pageref{firstpage}--\pageref{lastpage}} \pubyear{2010}

\maketitle

\label{firstpage}

\begin{abstract}
Current results from the \Lya forest assume that the primordial
power spectrum of density perturbations follows a simple power-law
form, with running. We present the first analysis of \Lya data to study
the effect of relaxing this strong assumption on primordial and astrophysical constraints. 
We perform a large suite of numerical simulations, using them to
calibrate a minimally parametric framework for describing the power spectrum.
Combined with cross-validation, a statistical technique   
which prevents over-fitting of the data, this framework allows us to 
reconstruct the power spectrum shape without strong prior assumptions. 
We find no evidence for deviation from scale-invariance; 
our analysis also shows that current \Lya data do not have 
sufficient statistical power to robustly probe the shape of the power spectrum at these scales. 
In contrast, the ongoing Baryon Oscillation Sky Survey will be able to do so with high precision. 
Furthermore, this near-future data will be able to 
break degeneracies between the power spectrum shape and astrophysical parameters. 
\end{abstract}

\begin{keywords}
cosmology: theory - methods: numerical - methods: statistical - galaxies: intergalactic medium
\end{keywords}

\section{Introduction}

The primordial power spectrum of density fluctuations underpins much of modern cosmology. On large scales, it
has been measured with high precision by cosmic microwave background (CMB) experiments (e.g. \citealt{WMAP7} and references within). 
In order to improve 
our knowledge of its scale-dependence, we turn to smaller scales, and astrophysical measurements probing later epochs in 
the evolution of the Universe. In this paper, we shall examine constraints from the data set which has probed the smallest 
scales to date: the \Lya forest.

The \Lya forest consists of a series of features in quasar spectra due to scattering of quasar photons with neutral hydrogen. 
Since hydrogen makes up most of the baryonic density of the Universe, the \Lya forest 
traces the intergalactic medium (IGM), and thus the baryonic power spectrum, on scales from 
a few up to tens of Mpc. This makes it the only currently 
available probe of fluctuations at small scales in a regime when the corresponding density fluctuations were 
still only mildly non-linear, thereby simplifying cosmological inferences. A number of authors have examined 
the cosmological constraints from the \Lya forest in the past 
(\citealt{Croft:1997, Theuns:1998, McDonald:2000, Hui:2001, Viel:2001, Gnedin:2002, McDonald:2004pk, Viel:2005,Lidz:2006}), 
while \cite{Seljak:2005, Seljak:2006} examined constraints combined with other data sets.
For a review of the physics of the IGM and its potential for cosmology, see \cite{Meiksin:2009}.

Previous analyses assumed that the primordial power spectrum on \Lya scales is described by a 
nearly scale-invariant power law -- a strong prior -- and proceeded with parameter estimation under this assumption. 
In contrast, in this work we attempt to constrain the shape and amplitude of the primordial power spectrum at these scales using 
minimal prior assumptions about its scale-dependence. 

In view of the observational effort dedicated to the \Lya forest, and its promise as a probe 
of the primordial power spectrum, in this work we shall explore the possibilities of going beyond parameter fitting.
To give us insight into the underlying model for the power spectrum shape, which parameter estimation by itself cannot do, our present application to \Lya data should therefore ideally assume full shape freedom throughout the analysis. As a nearly scale-invariant primordial power spectrum 
is a generic prediction of the simplest models of inflation, a minimally parametric reconstruction can be a powerful test of inflationary models.
\Lya constrains the smallest cosmological scales; thus, it provides the longest lever-arm when combined with the statistical power and robustness of CMB data, yielding the best opportunity presently available to understand the overall shape of the power spectrum. 

The main \Lya observable, the flux power spectrum, does not have a simple algebraic 
relationship to the matter power spectrum. By $z\sim3$, the absorbing structures are weakly nonlinear, 
and are also affected by baryonic physics. Hence, to establish the relationship between the primordial power spectrum and 
the flux power spectrum, we must resort to hydrodynamical simulations. 
The initial conditions used in our simulations allow for considerable freedom in the shape of the primordial power spectrum, and this allows us to recreate the \Lya forest resulting from generic power spectrum shapes. Using an ensemble of simulations which sample the parameter space required to describe the flux power spectrum, we construct a likelihood function which can be used to perform minimally parametric reconstruction of the primordial power spectrum, while simultaneously constraining parameters describing IGM physics. 

A statistical technique called cross-validation (CV) is used to robustly reconstruct the primordial power spectrum and Markov Chain Monte Carlo (MCMC) techniques are used to obtain the final constraints. The statistical approach parallels \cite{Sealfon:2005} and \cite{Verde:2008}, who applied the same method to data from the CMB and galaxy surveys. \cite{Peiris:2009} added the current \Lya forest data to the joint analysis with larger scale data, via the derived constraints on the small-scale matter power spectrum from \cite{McDonald:2004pk}. However, these latter constraints were derived assuming a tight prior on the shape of the primordial power spectrum at \Lya scales -- an assumption which we drop in this work. In our analysis we consider both the flux power spectrum determined by \cite{McDonald:2004data} from low-resolution quasar spectra obtained during the SDSS, and simulated data for the upcoming Baryon Oscillation Sky Survey (BOSS: \citealt{Schlegel:2009}).

This paper is organized as follows. In Section \ref{sec:methods} we review the framework for 
power spectrum reconstruction and describe the details of the simulations and parameter estimation setup.
Section \ref{sec:datasets} describes the data,
and results are presented in Section \ref{sec:results}. We conclude in Section \ref{sec:discussion}. 
Technical details of our calculations are relegated to Appendices \ref{ap:absorption} and \ref{ap:converge}.

\section{Methods}
\label{sec:methods}

In this section, we describe the statistical technique used in this paper, and how we built the likelihood function for minimally parametric reconstruction from \Lya data. Section \ref{sec:pkrecon} describes the framework for power spectrum reconstruction in general terms, while Section \ref{sec:knots} gives further details of our specific implementation of this framework. Sections \ref{sec:simulations} - \ref{sec:fluxpow} detail numerical methods used to extract a flux power spectrum from a given primordial power spectrum. Finally, Section \ref{sec:parameter} describes  the parameter estimation implementation.

\subsection{Power spectrum reconstruction}
\label{sec:pkrecon}

Previous analyses of the Lyman-$\alpha$ forest (\citealt{McDonald:2004pk, Viel:2004}) have assumed 
that the primordial power spectrum is a nearly scale-invariant power law, of the form

\begin{align}
        P(k) = A_\mathrm{s}\left(\frac{k}{k_0}\right)^{n_\mathrm{s}-1 + \alpha_\mathrm{s} \ln k},
        \label{eq:pk}
\end{align}
and then constrained $A_\mathrm{s}$,$n_\mathrm{s}$ and $\alpha_\mathrm{s}$. In this work we 
will follow the same spirit as \cite{Sealfon:2005, Verde:2008}, going
beyond parameter estimation in an attempt to deduce what the \Lya forest data can tell us 
about the shape of the power spectrum under minimal prior assumptions. A major challenge involved in all such reconstructions is to avoid 
over-fitting the data; it is of little use to produce a complex function that fits the data set extremely well if 
we are simply fitting statistical noise. Equally, an overly prescriptive function which is a poor 
fit to the data should be rejected. To achieve this balance, we add an extra term, $\mathcal{L}_P$, to the likelihood function 
which penalizes superfluous fluctuations. Schematically, the likelihood function is:
\begin{equation}
        \log \mathcal{L} = \log \mathcal{L}(\mathrm{Data} | P(k)) + \lambda \log \mathcal{L}_{\mathrm{P}}\,,
        \label{eq:likelihoodschematic}
\end{equation}
where the form of $\mathcal{L}_\mathrm{P}$ will be discussed shortly. Equation \ref{eq:likelihoodschematic} 
now contains an extra free parameter which measures the magnitude of the smoothing required; the penalty weight $\lambda$. 
As $\lambda \to \infty$ the likelihood will implement linear regression. For particularly clean data, carrying no evidence for any feature in the $P(k)$, $\lambda$ should be large.
Data carrying strong evidence for $P(k)$ features would be best analysed with a small value of $\lambda$.
We need a method of determining, from the data, the optimal penalty weight. Our chosen technique is called CV 
(\citealt{Green:1994}), which quantifies the idea that a correct reconstruction of the underlying information should accurately 
predict new, independent data. 

The variant of CV used in this paper splits the data into three sets. The function is reconstructed using two of these sets (training sets). 
The likelihood (excluding the penalty term) of this reconstruction, given only the data in the remaining set (validation set), is calculated. 
This step is called validation; because the data in each set are assumed to be independent, 
we now have a measure of the predictivity of the reconstruction. 
Validation is repeated using each set in turn and the total CV score is the sum of all three validation likelihoods. 
The optimal penalty is the one which maximizes the CV score. 

More generally, CV splits the data into $k$ independent sets, with 
$2 < k \leq n$, where $n$ is the number of data points. $k-1$ sets are used for training, and the remaining set for validation. 
Larger $k$ allows for a bigger training set, and thus better estimation of the function to be validated against, but for most practical problems large $k$ 
is computationally intractable. We have chosen $k=3$ as a compromise. We verified that using 
$k=2$, following \cite{Verde:2008}, made a negligible difference to our results despite the smaller training set size.

CV assumes that each set is uncorrelated;
a mild violation of this assumption will lead to an underestimation of errors, but not a systematic bias 
in the derived parameters \cite{Carmack:2009}. Our data include a full covariance matrix, and so we are able 
to verify that correlations between the sets are weak.

The minimally parametric framework applied in this paper follows that of \cite{Sealfon:2005, Verde:2008} and \cite{Peiris:2009}. 
It uses cubic splines to reconstruct a function $f(x)$ 
from measurements at a series of points, $x_i$, called the knots. The function value between each pair of knots 
is interpolated using a piecewise cubic polynomial. The spline is fully specified by the knots, 
continuity of the first and second derivatives, and boundary conditions on the second derivatives at the exterior knots
(the knots at either end of the spline). The splines have vanishing second derivative at the exterior knots. 
If the power spectrum is given by smoothed splines, the form of the likelihood function given above is
\begin{align}
        \log \mathcal{L} = \log \mathcal{L}(\mathrm{Data} | P(k)) &+ \lambda \int_k d \ln k (P''(k))^2,\\ 
        \mathrm{where}\; P''(k) &= \frac{d^2 P}{d(\ln k)^2} \nonumber.
        \label{eq:penalty}
\end{align}

\subsection{Knot placement}
\label{sec:knots}

The number and placement of the knots is chosen initially and kept fixed throughout the analysis. Once 
there are sufficient knots to allow a good fit to the data, adding more will not alter the shape of the 
reconstructed function significantly. In choosing the number of knots, we seek to find a balance between 
allowing sufficient freedom in the power spectrum, and having few enough parameters that the data are still able to provide 
meaningful constraints on two sets out of three when subdividing the data into the training and validation sets, as described above. 
Available computing resources limit us in any case to considering only a few knots.
We fit the primordial power spectrum with a four-knot spline for the \Lya forest $k$-range. 
The flux power spectrum is available in twelve $k$-bins, so there are 
three bins per knot, which should allow sufficient freedom. By comparison, \cite{Peiris:2009} used 
seven knots to cover the $k$-range spanned by CMB, galaxy surveys and \Lya data, with a 
single knot for the \Lya forest.

The SDSS flux power spectrum covers the range of scales, in velocity units, of $k_v = 1.41\times 10^{-3} - 0.018\, \mathrm{s/km}$. 
Dividing by a factor of $H(z)/(1+z)$ converts to comoving distance coordinates, so the constraints on 
the matter power spectrum are on scales of roughly $k = 0.4 - 3\ \Mpc^{-1}$. 
In this range of scales we place four knots (A--D, from large to small scales) evenly in log space. Numerical details of the knots 
are shown in Table \ref{tab:knots}. The maximum and minimum values of $P(k)$ given there for each knot are simply the extremal 
values covered by our simulations. Simulation coverage of $P(k)$ has been expanded where necessary to fully cover the range 
allowed by the data.

\begin{table}
\begin{center}
\begin{tabular}{|c|c|c|c|}
\hline
Knot & Position          &\multicolumn{2}{|c|}{$P(k)$ ($/ 10^{-9}$)} \\ 
&  ($\Mpc^{-1}$)   & Minimum & Maximum \\
\hline
A  &  0.475           & $0.83$ &  $3.25$ \\
B  &  0.75            & $0.60$ &  $3.23$ \\
C  &  1.19            & $0.60$ &  $3.67$ \\
D  &  1.89            & $0.53$ &  $4.16$ \\
\hline    
\end{tabular}
\end{center} 
\caption{Positions of the knots. The maximum and minimum values of $P(k)$ are
the extremal values covered by our simulations.
Fixed knots are not shown, but are discussed in the text.}
\label{tab:knots}
\end{table}

We must specify the primordial power spectrum on scales well outside the range probed by data, even though they have no effect 
on the \Lya forest. This is for two reasons. The first is that
when running a simulation we must have a well-defined way to perturb the initial particle grid for all scales included in the simulation.
In order to ensure that the scales on which we have data are properly resolved, we also need to simulate larger and smaller scales
, and these require a defined power spectrum. The second reason is that our interpolation scheme works best when 
the perturbations induced by altering one of the knots are reasonably local. Adding extra end knots helps to prevent large secondary 
boundary effects, which would make interpolation far more difficult. 

For numerical stability reasons, we would like the amplitude of fluctuations on 
these scales to be reasonably constant, but do not wish to make strong assumptions about the amplitude of the power spectrum there. 
Therefore we add two ``follower'' knots at each end of the spline. The amplitude is fixed to follow the nearest parameter knot, assuming 
that between follower and followed, the shape is a power law with $n_s = 0.97$.
\footnote{Hence, if the amplitude of the power spectrum at the D-knot is $P^D$, the power spectrum at the follower knot
has the amplitude $P^D (k^D/k^{D+1})^{0.03}$, where $k^D$ is the position of the D knot and $k^{D+1}$ the position of the follower.}
The two follower knots are at scales of $k= (0.15, 4)\, \Mpc^{-1}$. 

\begin{figure}
\centering
\includegraphics[width=0.9\columnwidth]{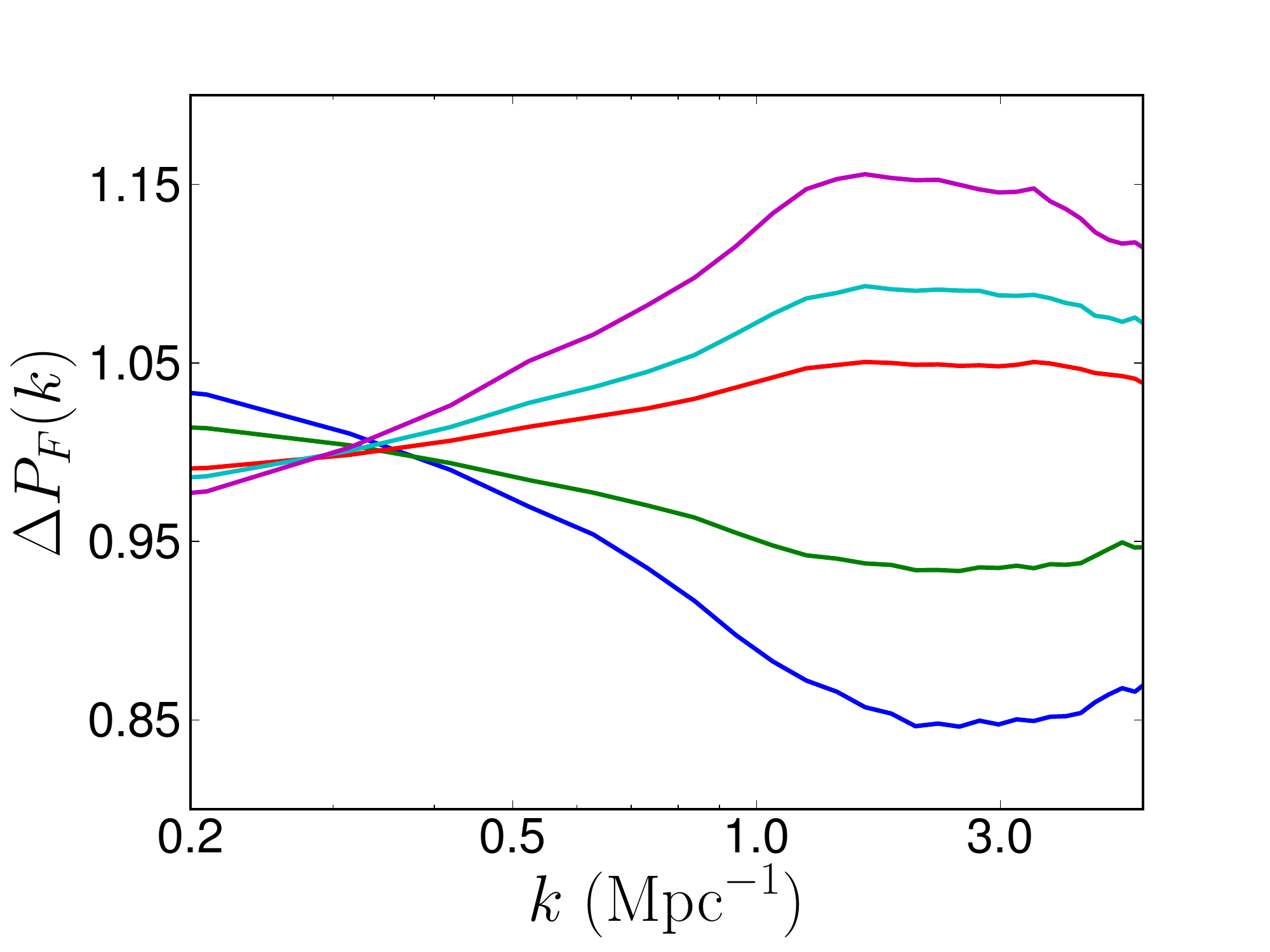}
\caption{Effect on the flux power spectrum of varying the D knot at $z=3$. On a scale where the best-fitting amplitude is $0.9$, the amplitudes of the D knot are, from the lowest line upwards, $0.5$, $0.7$, $1.1$, $1.3$ and $1.7$. Non-linear growth tends to erase dependence on the initial conditions, so the effect is smaller at lower redshifts.}
\label{fig:powerspec}
\end{figure}

We also add a few knots, even more distant from the scales probed by the data, with completely fixed amplitudes consistent 
with the WMAP best-fitting power spectrum. The amplitude of the primordial power spectrum on these scales does not significantly
affect results; we have added knots here so that the initial density field is well-defined on a larger range of scales than probed by the 
simulation. This allows us to avoid any boundary effects associated with the ends of the spline.
These fixed knots are at $k = (0.07, 25,40)\ \Mpc^{-1}$, with amplitudes of $(2.43,2.03, 2.01)\ \times 10^{-9}$. 
Fig. \ref{fig:powerspec} shows the effect of altering the amplitude of the D knot on the flux power spectrum. 

\subsection{Simulations}
\label{sec:simulations}

In this study, full hydrodynamical simulations were run using the parallel TreePM code 
{\small GADGET}-$2$ (\citealt{Springel:2005}). {\small GADGET} computes long-range gravitational forces using a particle grid, 
while the short-range physics are calculated using a particle tree. 
Hydrodynamical effects are included by having two separate particle types; 
dark matter, affected only by gravity, and baryons, modelled using 
smoothed particle hydrodynamics (SPH), where particles are supposed to approximate density elements in the matter fluid. 
The rest of this section gives technical details of our simulations and the included astrophysics, 
and may be skipped by the reader interested only in the cosmological implications. 

{\small GADGET} has been modified to compute the ionisation state of the gas using radiative cooling and ionisation physics as originally 
described by \cite{Katz:1995}, and used in \cite{Viel:2005}. Star formation is included via a simplified prescription 
which greatly increases the speed of the simulations, where all baryonic particles with overdensity $\rho/\rho_0 > 10^3$ 
and temperature $T < 10^5 K$ are immediately made collisionless. \cite{Viel:2004} compared simulations using this prescription
with identical simulations using a multi-phase model, and found negligible difference in the \Lya statistics. Additionally, all 
feedback options have been disabled, and galactic winds neglected; \cite{Bolton:2008} found
that winds have a small effect on the \Lya forest. 
The gravitational softening length was set to $1/30$ of the mean linear inter-particle spacing.

The gas is assumed to be ionised by an externally specified, spatially homogeneous UV background, 
based on the galaxy and quasar emission model of \cite{HaardtMadau}. We follow previous analyses in assuming
that the gas temperature is initially in equilibrium with the CMB, that the gas is in ionisation equilibrium, optically thin, 
and that we can neglect metals and evolution of elemental abundances. \Lya absorption arises largely from near mean-density 
hydrogen, 
which should undergo little chemical evolution, so using a simplified star formation criterion and neglecting metals is 
physically well-motivated. Assuming that the gas is optically thin and in ionisation equilibrium will break down during reionisation, 
but at the redshifts we are interested in, we can model the effect of non-instantaneous reionisation by increasing the 
photo-heating rate, as described in \cite{Viel:2005}. 

The fiducial simulation for this paper has a box size of $60\ \Mpch$ and $2 \times 400^3$ gas and dark matter particles, 
[which we will write as ($60$, $400$) in future], and runs from
$z=199$ to $z=2$. Snapshots are output at regular intervals between redshift $4.2$ and $2.0$. Initial conditions were generated using N-GenICs, 
modified to specify the primordial power spectrum by a spline, and use separate transfer functions for baryons and dark matter, 
as calculated using CAMB (\citealt{camb}).

For knots B and C, we used the above fiducial parameters for box size and particle resolution. For the D knot, we slightly compromised 
on box size in favour of particle resolution, and used simulations of ($48$, $400$), since we found that the D knot had a negligible affect 
on the largest scales. To fully capture the behaviour of the A knot, we used larger simulations with ($120$, $400$). 
We have used different sized simulations to ensure that for each knot, the characteristic scales representing it have very good numerical 
convergence; this issue is addressed in full in Appendix \ref{ap:converge}. 
Our ability to do this is one technical advantage of our approach compared with previous studies; if we were to alter the amplitude of the
whole power spectrum, we would need to achieve convergence over all the relevant scales at once. In our approach, each simulation only 
needs strict convergence over the narrow range of scales probed by a single knot.

\subsection{IGM thermodynamics}
\label{sec:thermo}

Constraints on the thermal history of the IGM are given in terms of the parameters of a polytropic temperature-density relation
\begin{equation}
        T = T_0 \left(\frac{\rho}{\rho_0}\right)^{\gamma-1}\,,
        \label{eq:igmeos}
\end{equation}
where a given SPH particle has temperature $T$ and density $\rho$. $T_0$ and $\rho_0$ are the average quantities for 
the whole simulation snapshot.
To determine $T_0$ and $\gamma$ from a simulation box, a least-squares fit is performed from low-density
particles satisfying
\begin{equation}
        -1.0 < \log\left( \frac{\rho}{\rho_0} \right) < 0\, .
        \label{eq:T0}
\end{equation}
Regions that are less dense than the lower limit above are ignored because they are poorly resolved in SPH simulations (\citealt{Bolton:2009}). 
The simplified star formation criterion 
means that many overdensities have been turned into stars, and their baryonic evolution not followed; hence they are also neglected. 
Both $\gamma$ and $T_0$ are assumed to follow a power law broken at $z=3$ by HeII reionisation (\citealt{Schaye:1999}), so that they are given by:

\begin{align}
        \gamma =& \begin{cases}
                \gamma^A \left[(1+z)/4\right]^{d\gamma^S}&       \text{if $z<3$}, \\
                \gamma^A \left[ (1+z)/4 \right]^{d\gamma^R}&     \text{if $z>3$}.
        \end{cases} \\
        T_0 =& \begin{cases}
                T_0^A \left[(1+z)/4\right]^{dT_0^S}&      \text{if $z<3$}, \\
                T_0^A \left[ (1+z)/4 \right]^{dT_0^R}&        \text{if $z>3$}.
        \end{cases}
        \label{eq:thermalhistory}
\end{align}
When performing parameter estimation, we marginalize over $\gamma^A,\, T_0^A$ and $d\gamma^{S,R},\, dT_0^{S,R}$. The different thermal histories 
were constructed by modifying the fiducial simulation's photo-heating rate as described in Section $2.2$ of \cite{Bolton:2008}. 

The effective optical depth is described by a power law, with parameters:
\begin{equation}
        \tau_\mathrm{eff} = \tau^A_\mathrm{eff} \left[ (1+z)/4 \right]^{\tau^S_\mathrm{eff}}\, .
        \label{eq:taueff}
\end{equation}

Previous studies (\citealt{McDonald:2004pk, Viel:2005}) used the same transfer function for both dark matter and baryon particles;
we have used different transfer functions for baryon and 
dark matter species. At our starting redshifts, the transfer functions for the baryons are about $10\%$ 
lower than for the dark matter on these scales, because baryon fluctuations have not grown as fast 
during tight coupling. Once they have decoupled from the photons, the baryons fall 
into the potential wells of the dark matter, and by $z=1$, the linear transfer functions
are almost identical. 
At redshifts $2-3$, however, the effect is small but noticeable, and accounts for a $2\%$ scale-independent 
drop in the power spectrum. This is too small to affect current data, but could be potentially important for analysing BOSS data. 

\subsection{The flux power spectrum}
\label{sec:fluxpow}

In the case of Lyman-$\alpha$, the observable is not a direct measurement of the 
clustering properties of tracer objects, as in galaxy clustering, but the 
statistics of absorption along a number of quasar sightlines. 
Therefore we define the flux, $\mathcal{F}$, as
\begin{equation}
        \mathcal{F} = \exp (-\tau),
        \label{eq:flux}
\end{equation}
where $\tau$ is the optical depth. We define the flux power spectrum as
\begin{align}
        P_F(k) &= | \tilde{\delta_F}(k) |^2, \nonumber\\
        \delta_F &= \frac{\mathcal{F}}{\bar{\mathcal{F}}} - 1 \,.
        \label{eq:fluxpk}
\end{align}
Here $\bar{\mathcal{F}}$ is the mean flux. 
The tilde denotes a Fourier transformed quantity, where our Fourier conventions, used throughout, are:
\begin{equation}
        \tilde{f}(k) = \int f(x) e^{i kx} \mathrm{dx}  \,.
        \label{eq:fourier}
\end{equation}

To aid the eventual understanding of our results, we digress slightly here to review the physical effects of the various thermal parameters on 
the flux power spectrum.
The mean flux, essentially a measure of the average density of neutral hydrogen, has a large impact on the amplitude of the flux power spectrum. 
Cosmological information from the \Lya forest is obtained through examining the power spectrum shape and its redshift dependence. 
The effect of a higher temperature, as preferred by the flux power spectrum, is to suppress power predominantly on small scales, 
as a higher temperature wipes out small-scale structure in the baryons. The exponent of the temperature-density relation, 
$\gamma$, controls the temperature difference between voids and overdensities. A higher $\gamma$ makes voids cooler and overdensities 
hotter. At high redshifts, where much of the \Lya absorption comes
from voids, the effect of an increased $\gamma$ is to decrease the temperature of the \Lya emitting regions, so there is relatively more 
small-scale structure. At low redshifts, however, most of the \Lya absorption comes from near mean density material, and so an increased
$\gamma$ increases the temperature, decreasing the amount of small-scale structure. For further details of the physical effects of the
various parameters, see Section 4.2.1 and Fig. 3 of \cite{Viel:2005}, as well as Fig. 11-13 of \cite{McDonald:2004pk}.

Current constraints on $P_F$ are given by \cite{McDonald:2004data}, determined 
from $\sim 3000$ SDSS quasar spectra at $z=2-4$. 

Each simulation snapshot was processed to 
generate an averaged flux power spectrum as follows. 
First, $8000$ randomly placed simulated quasar sightlines were drawn through 
the simulation box. For a $60\ \Mpch$ box, this constitutes an average spacing between 
sightlines of $670\ h^{-1}\mathrm{kpc}$, corresponding to scales of roughly $k=10\ \Mpch$, 
far smaller than the scales probed by the \Lya forest. We verified that doubling the number of sightlines 
to $16000$ made a negligible difference to the resulting power spectra.

When calculating absorption, particle peculiar velocities were included, 
which increases the (non-rescaled) magnitude of the power spectrum by approximately $10\%$. 

To generate the flux power spectrum, the absorption due to each SPH particle near the sightline is calculated, giving us a number of simulated quasar 
spectra, which are smoothed with a simple boxcar average. Each spectrum is rescaled by a constant so that the mean 
flux across all spectra and absorption bins matches that observed by \cite{Kim:2007}. This rescaling hides our ignorance
of the amplitude of the photo-ionizing UV background.
The mean over all the rescaled spectra is then used as the extracted flux power spectrum for the box. For further details of how we computed the 
absorption, see Appendix \ref{ap:absorption}.

We follow previous work in not attempting to model continuum fitting errors. 
The Si III contamination found by \cite{McDonald:2004data} is modelled by assuming a linear bias correction 
of the form $P_F' = \left[(1+a^2)+2a \cos ( v k)\right] P_F$, with 
$a = f_{\mathrm{SiIII}}/(1-\bar{\mathcal{F}})$, $f_{\mathrm{SiIII}} = 0.011$, and $v=2271\, \mathrm{km}/\mathrm{s}$.

Finally, since high-density, damped \Lya systems (DLAs) are not modelled by our simulations, 
we add a correction to the flux power spectrum to account for them, of the form calculated by \cite{McDonald:2004dla}. 
The amplitude of this correction is a free parameter, and will be discussed further in Section \ref{sec:mcmcparam}.

We checked the convergence of our simulations with respect to box size and particle resolution.
Here we give only a brief summary of the results; 
further details may be found in Appendix \ref{ap:converge}.
For the highest redshift bins at $z=4.2$, $4.0$ and $3.8$, increasing the particle resolution had a large effect on the 
flux power spectrum. Achieving numerical convergence for the \Lya forest at high redshift
is challenging, because most of the signal for the \Lya forest is coming from poorly resolved underdense regions.
In addition, current data at high redshifts are much more noisy than at low redshifts, 
and future surveys will not probe these redshifts at all. Accordingly, we follow \cite{Viel:2005}
and do not use the three highest redshift bins in our analysis. 

At lower redshifts, and except in the smallest and largest $k$-bins, the change with increased particle resolution was small.
On the smallest scales, however, there was a change of around $5\%$ in each bin. This increase is systematic, and so we correct for it as described 
in Appendix \ref{ap:converge}. The larger box increased power on the largest scales by around $5\%$, 
due to sample variance in the simulation box. The methodology we 
used to correct for this effect is again detailed in Appendix \ref{ap:converge}. 

The above figures were the dominant errors in our modelling of the flux power spectrum. Uncorrected modelling errors 
are therefore $\lesssim 2\%$ of the flux power spectrum in each bin, far below the current measurement error of $\sim 12\%$ in each bin of the flux power spectrum, and on the order of the expected statistical errors for the BOSS survey, which are $\sim 1.5\%$. A significant decrease 
in modelling errors would require the use of simulations with improved particle resolution, which are beyond
the computational resources available to us.

\subsection{Parameter estimation}
\label{sec:parameter}

So far we have given a formula for the primordial power spectrum, and described how we use it to extract 
a flux power spectrum to compare with observational data. In this section, we shall describe how we 
actually performed that comparison. First we describe a scheme for robustly interpolating the parameter space to obtain flux power spectra corresponding to parameter combinations which we have not simulated, following \cite{Viel:2005}. Secondly, we 
describe the parameters of the Monte Carlo Markov chains we used for parameter estimation. For more details of MCMC, 
see, for example \cite{cosmomc}. 

\subsubsection{Parameter interpolation}
\label{sec:interp}

Directly calculating a flux power spectrum from a given set of primordial fluctuations 
requires a hydrodynamical simulation. This makes it impractical to directly calculate 
$P_F$ for every possible set of input parameters. 
Instead, simulations are run for a representative sample and other results are obtained from these via interpolation. 
We assume that the flux power spectrum varies smoothly around the best-fitting model, parametrize this variation 
with a quadratic polynomial for each data point, and then check that this accurately predicts new points. 
If we have some simulation with a parameter vector which differs from a `best-guess' simulation by $\delta p_i$, 
the corresponding change in the flux power spectrum, $\delta P_F$, is given by
\begin{equation}
        \delta P_F =  \sum_j \alpha_j \delta p_j+ \beta_j \delta p_j^2 \,.
        \label{eq:fluxpower}
\end{equation}
The coefficients of this polynomial are constrained by performing a least-squares fit 
to flux power spectra generated by numerical simulations. We experimented with 
including cross-terms (of the form $p_i p_j$), but found that this did not significantly
improve the accuracy of the interpolation.

\begin{figure}
\includegraphics[width=0.9\columnwidth]{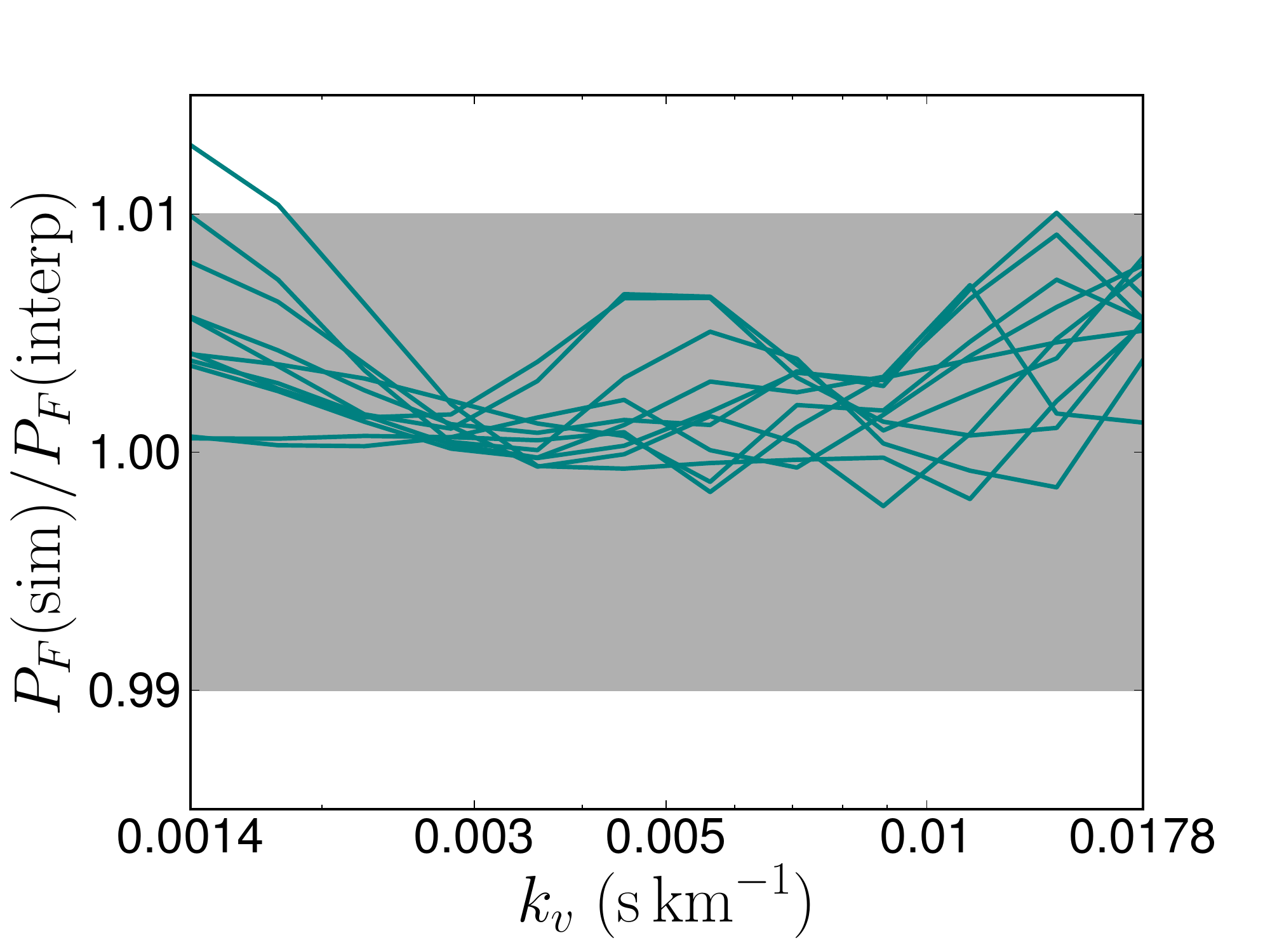}
\caption{The difference between the flux power spectrum as obtained from interpolation, and directly by simulation. Here only the C and D knots have been changed from their initial values.
Each line represents simulation output at a different redshift bin, between $z=2.0$ and $z=4.2$. The grey band shows $1\%$ error bars.
}\label{con:intp}
\end{figure}

To estimate the interpolation coefficients, we used seven simulations for each of our four power spectrum parameters, one of which 
was used to test the accuracy of the interpolation. 
To check for correlation between parameters, we simulated varying two neighbouring knots at once.
As the greatest effect of each knot on the flux power spectrum is over a localized range of scales, our interpolation errors should be maximal here.
We needed only four simulations per thermal history parameter, and checked we could accurately predict 
$\delta P_F$ for a very different thermal history.
As a final interpolation verification, we performed a simulation where all six parameters were changed simultaneously.
Fig. \ref{con:intp} shows the interpolation errors for one of our tests, which are around $1\%$ of the total change for each bin.
This is smaller than the expected statistical errors for BOSS, and was replicated by our other test simulations.

\subsubsection{MCMC methodology}
\label{sec:mcmcparam}

To perform parameter estimation, we use a version of the publicly available CosmoMC \cite{cosmomc} code, with a modified likelihood function 
as described in Section \ref{sec:pkrecon}. 

We marginalize over four parameters for the four knots, with priors as specified in Table \ref{tab:knots}, and over eight parameters
of the thermal history, as described in Section \ref{sec:simulations}. 

We follow the advice of \cite{McDonald:2004data}, and add a number of nuisance parameters to the SDSS data, all with Gaussian priors.
To parametrize uncertainty in the resolution of the spectra, we add a parameter $\alpha^2$ with prior $0\pm 49$, and multiply the 
flux power spectrum by $\exp \left( -k^2 \alpha^2 \right)$. The effect of an increased $\alpha^2$ is 
therefore to damp power on the very smallest scales. 
Each redshift bin has one parameter, $f_i$, to describe uncertainty in the subtraction of background noise, with a prior of $0 \pm 0.05$. 
To marginalize over the uncertainty in the effect of DLAs, we add $A_{\mathrm{damp}}$, with a prior of $1 \pm 0.3$. 
The effect of this correction is to increase the slope of the flux power spectrum.

We also marginalize over residual uncertainties in the Hubble parameter, $h$ and $\Omega_M$, using flat priors of 
$0.2 < \Omega_M < 0.4$, and $0.5 < h < 0.9$. For the rest of our background cosmology, 
we assume parameters in agreement with those preferred by WMAP 7, including negligible 
gravitational waves and spatial curvature. The priors on $h$ and $\Omega_M$ make a negligible difference to our results, because 
both these parameters only weakly affect the \Lya forest. We assume $T_0 < 50000$ K and $0 < \gamma < 5/3$ on physical grounds; 
the temperature-density relation of the IGM cannot be steeper than the perfect gas law,
and very high temperatures would 
contradict independent measurements of the IGM temperature by \cite{Schaye:1999}.

\subsubsection{Cross-validation methodology}
\label{sec:crossvalid}

Cross-validation (CV) requires the splitting of the data set into $n$ independent sets. For best results, 
these sets should be as uncorrelated as possible. We choose to use alternating bins in $k$ for each set. 
For data with $n$ $k$ bins, the first set would consist of bins $1,4,7\dots$, the second bins $2,5,6\dots$ and 
the third similarly. 

To calculate the CV score, we estimate the best-fit from the two training sets, using an MCMC. The CV score 
for the remaining, validation, set is the likelihood of this best-fit. The total CV score for a given penalty 
is the sum of the CV scores for each set. 

\section{Datasets}
\label{sec:datasets}

\subsection{Current data from SDSS}
\label{sec:sdssdata}

The SDSS data used in this study consist of a best-fitting flux power spectrum
in $12$ $k$-bins and $11$ redshift bins, together with a covariance matrix and a set of vectors describing the 
foreground noise subtraction. It was analysed by \cite{McDonald:2004data}, and comes from $3000$ quasar spectra. 
Of these, $\sim 2000$ are at redshift $2.2-3$, and $\sim 1000$ above that. We use the $8$ redshift bins at $z<3.8$ only. 

We have chosen not to include any additional small-scale information based on high-resolution quasar spectra. 
In principle, this can help break degeneracies and should be included in future analyses. 
Currently, however, systematic error from such data sets is hard to quantify, and the optimal method for extracting the 
thermal state of the IGM is not yet clear. Our focus in this work has been robustness, and so we have 
limited ourselves to a single data set, whose properties have been extensively studied and 
are relatively well-understood.
\subsection{Simulated data from BOSS}
\label{sec:bosssimulate}

In this section we will describe our simulated data for forecasting constraints from BOSS, 
an ongoing-future survey which will acquire $1.6\times 10^5$ quasar spectra (\citealt{Schlegel:2009}), between $ z= 2.2 - 3.0$.
We need to simulate both a covariance matrix and a flux power spectrum. 

We have assumed that the noise per spectrum of the BOSS data 
will be approximately the same as they were for SDSS. This is a simple assumption, but broadly justified because both 
surveys use similar instruments (\citealt{Schlegel:2009}). Truly accurate modelling of the covariance matrix is 
impossible until the release of the final data, however, we expect our modelling of the BOSS covariance matrix 
to be completely adequate for a forecast.
Our simulated BOSS covariance matrix is simply the SDSS covariance matrix scaled to account for the increase in 
statistical power resulting from the much greater number of quasar sightlines. There are roughly $2000$ quasar sightlines 
in the SDSS sample below $z=3$, so the scale factor is $2000/160000 = 1/80$. 

To generate the flux power spectrum, we used cosmological parameters consistent with the best-fitting
results from WMAP 7, and thermal parameters consistent with theoretical expectations:
$\gamma \sim 1.45$ and $T_0 = 2.3\times10^3 \left[(1+z)/4\right]^{0.2}\,$K.
The effective optical depth was $\tau = 0.36 \left[(1+z)/4\right]^{3.65}$. 
The power spectrum amplitude was selected to match a spectrum with $\sigma_8 =0.8$ and $n_s =0.96$. 

We then added uncorrelated Gaussian noise with a variance given by the diagonal elements of the simulated BOSS covariance matrix. 
As BOSS will only take data at $z\leq 3$, we dispense with the thermal parameters for higher redshifts. 
The foreground noise properties of the BOSS data are expected to be similar to those of the SDSS data; we therefore 
leave the priors on the parameters measuring uncertainty in the noise subtraction and the parameter measuring resolution 
uncertainty, $\alpha^2$, unchanged. 

BOSS is also expected to determine the transverse flux power spectrum. Simulating the larger 
scales needed to properly model the effect of this is beyond the scope of this paper, and we refer the interested reader to
\cite{Slosar:2009, White:2010}.

\section{Results}
\label{sec:results}
\subsection{Current constraints}
\label{sec:sdssresults}

\begin{figure*}
\centering
\includegraphics[width=0.45\textwidth]{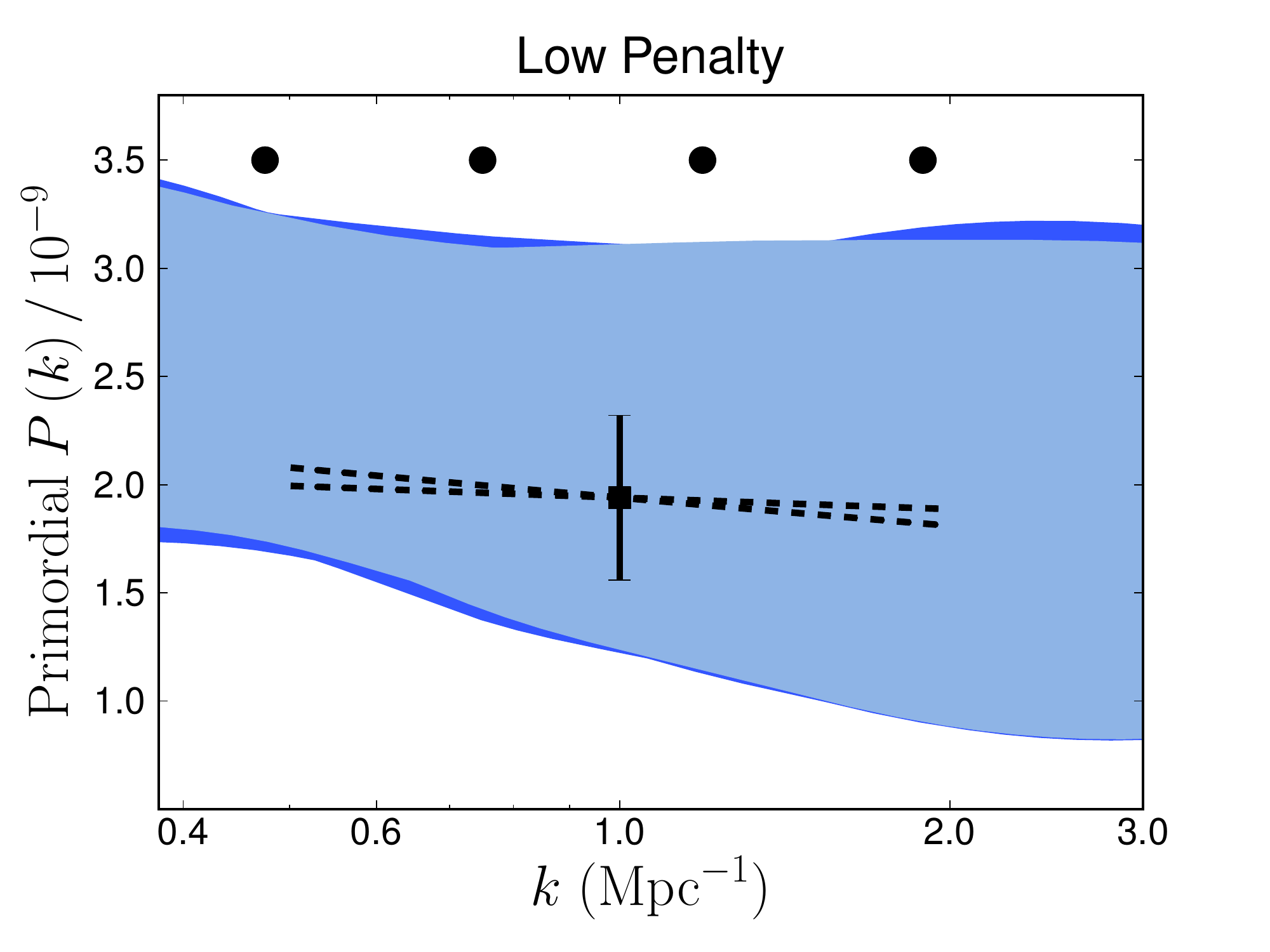} 
\includegraphics[width=0.45\textwidth]{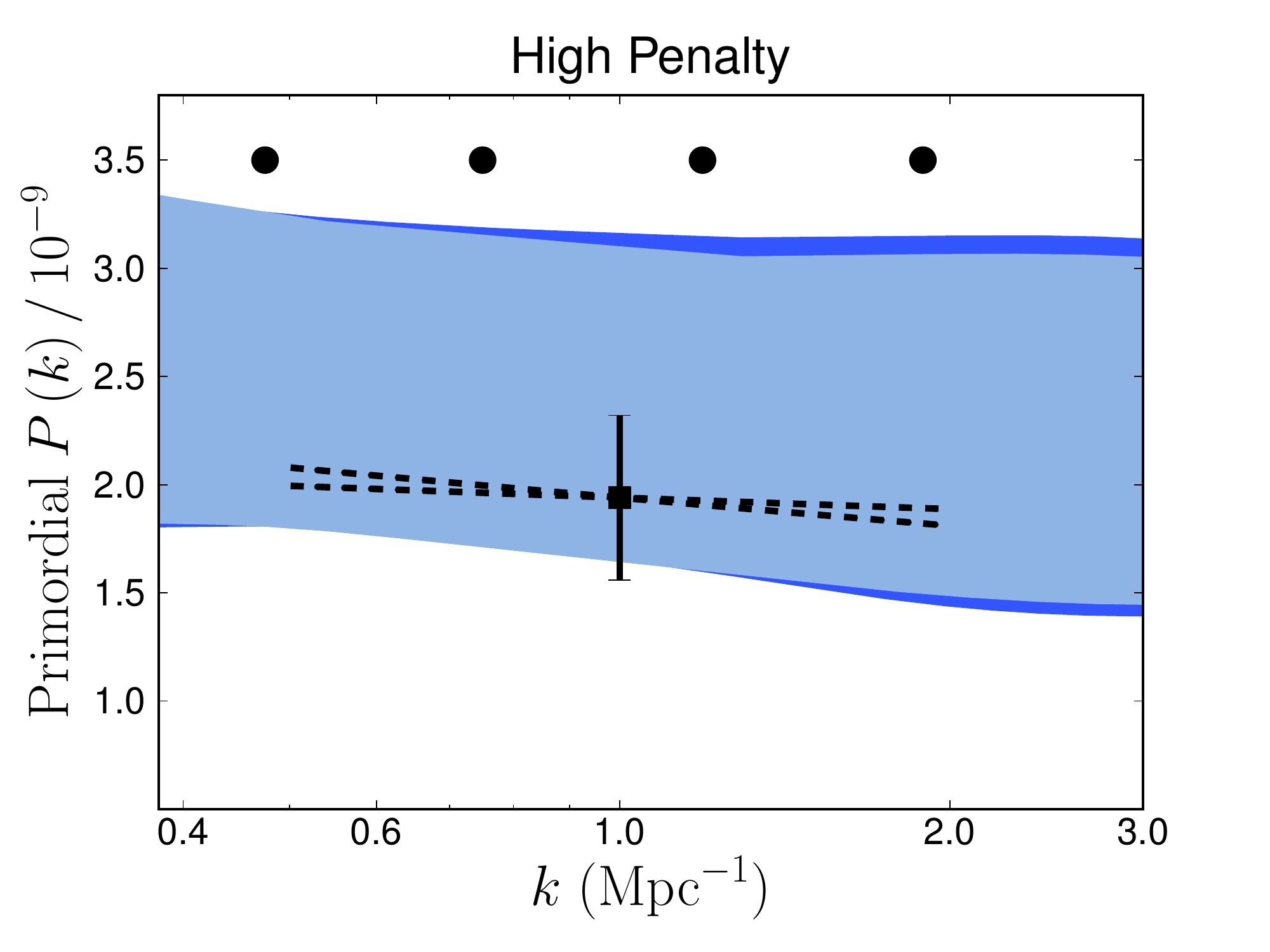}
\caption{Constraints on the primordial power spectrum from SDSS data from CV, 
for low (left) and high (right) penalties. 
Black circles show the positions of the knots, with arbitrary normalisation. 
The light blue regions show the top $68\%$ of likelihoods for SDSS data, 
while the dark blue regions show the $95\%$ likelihood range.
The black error bar shows the results of previous analyses (\protect\citealt{Viel:2009}) assuming a power-law power spectrum at $k=1\, \Mpc^{-1}$. 
The dashed lines show limits on the slope from that work.}\label{fig:current}
\end{figure*}

Fig. \ref{fig:current} shows the CV-reconstruction of the primordial power spectrum from SDSS data. We have emulated 
confidence limits by plotting the envelope of samples which have a likelihood in the top $68\%$ and the top $95\%$. At the $95\%$ level,
the power spectrum is allowed to oscillate more within the allowed envelope, but the size of the overall constraint on the amplitude does not 
greatly change, as found by \cite{Verde:2008}. 

We have shown plots for two penalties: one high, one low. This was because we have been unable to determine an optimal penalty from current data;
the CV score shows no significant variation, even when the penalty is having negligible impact on the likelihood. 
We interpret this to mean that the shape constraints on the primordial power spectrum from current Lyman-$\alpha$ data are very weak.
 
Previous analyses assumed a power-law prior for the shape of the primordial power spectrum, and constrained this slope and the overall 
normalisation from the same data used above. While such parameter estimation leads to tight constraints from the data (assuming the underlying 
shape prior is correct), relaxing this tight prior leads to the loss of ability to constrain the scale-dependent shape of the power spectrum. 
The current data can still be used as part of a minimally parametric primordial power spectrum if one exploits the extended range in scales 
that can be probed in combination with other data sets (\citealt{Peiris:2009}).

The black error bar in Fig. \ref{fig:current} shows a comparison with \cite{Viel:2009}. Our method gives results for the amplitude of the 
primordial power spectrum at \Lya scales which are completely consistent with that work, but somewhat weaker. This is to be expected; 
we are removing a tight prior on the shape of the power spectrum. For a very high penalty, i.e. the limit at which the implicit prior 
in our analysis approaches a power-law spectrum, we can reproduce the error bars of \cite{Viel:2009}. We are also in agreement with the results 
of an earlier analysis of the \Lya forest, \cite{McDonald:2004pk}, which constrained $\sigma_8 = 0.85 \pm 0.13$.

The corresponding constraints on $n_s$ from our reconstruction are extremely weak, especially for the low penalty: $n_s \sim 0.2 - 1.2$. 
The constraints on $n_s$ in \cite{Viel:2009}, in addition to the power-law prior, were greatly assisted by the 
fact that the pivot scale $k_0$ in Eq.~\ref{eq:pk} was chosen to be $k_0=0.002\, \mathrm{Mpc}^{-1}$; a small change 
in the slope of the power spectrum at $k_0$ leads to a large change in power spectrum amplitude 
by $k=1\, \mathrm{Mpc}^{-1}$. Here, we are trying to constrain a scale-dependent $n_s(k) = 1+\frac{d \ln P }{d \ln k}$ using only
the interval of scales sampled by \Lya forest. We find that, while the current \Lya data are able to constrain the 
amplitude of the power spectrum at these scales, they are 
not powerful enough on their own to significantly constrain the shape of the spectrum in a robust manner.
At no penalty do we see any evidence in the current data against a scale-invariant power spectrum.

\begin{figure}
\centering
\includegraphics[width=0.4\textwidth]{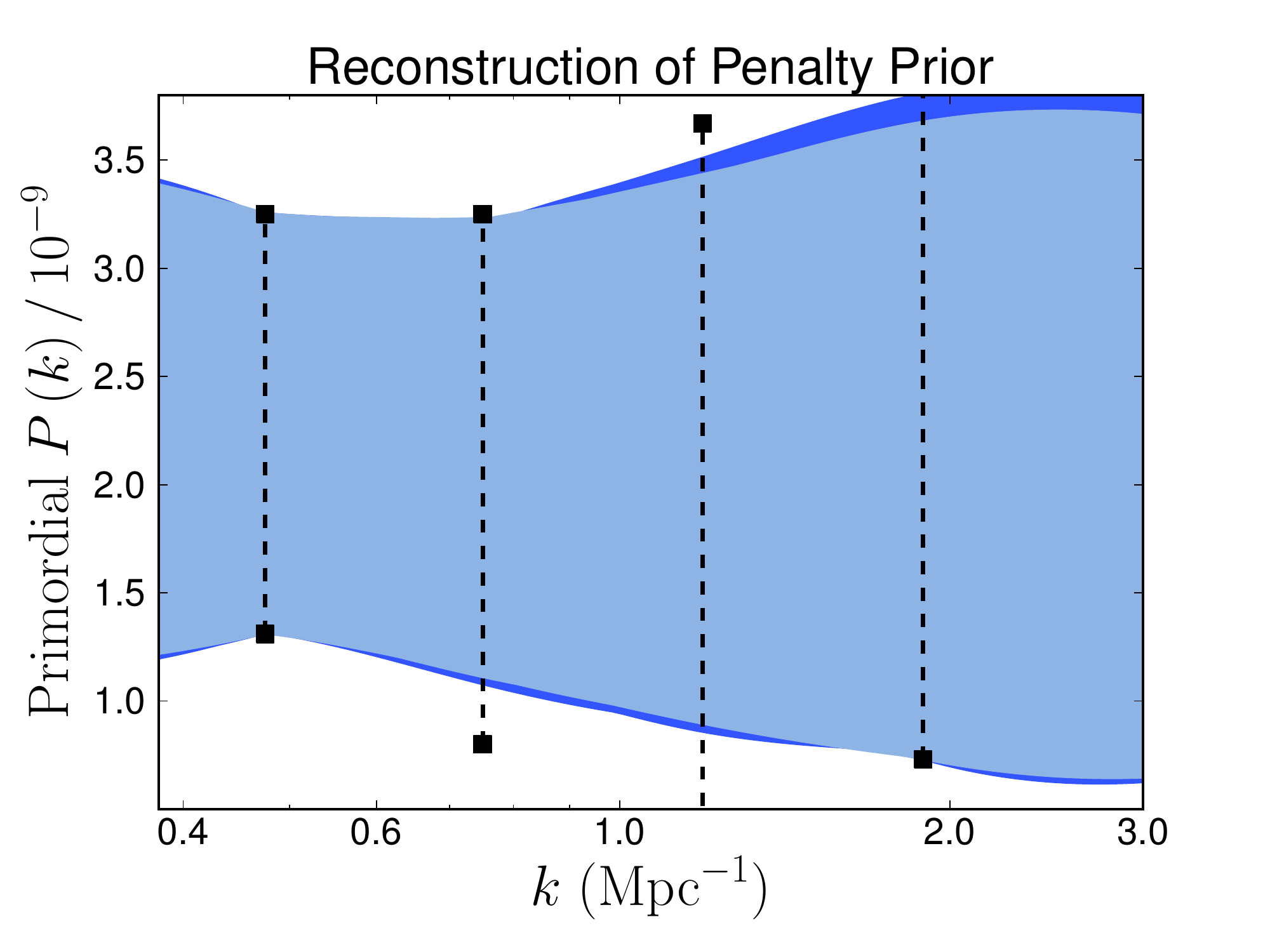}
\caption{Constraints on the primordial power spectrum from the penalty term alone, using the value in the ``low penalty'' plot of 
Fig. \protect\ref{fig:current} above. Dashed lines show the power specrum range sampled by the simulations. 
}\label{fig:prior}
\end{figure}

We can explicitly demonstrate that the current \Lya data add little information to a weak prior on the shape of the power spectrum in the following way. 
Fig. \ref{fig:prior} shows a minimally parametric reconstruction assuming the penalty designated ``low''.  These constraints were 
generated {\sl without using any data whatsoever}, and are similar to those obtained with the \Lya forest data. This figure shows clearly 
that our SDSS constraints are affected by the prior even for the low penalty. 
Since the penalty is proportional to $P''(k)$, it cannot determine the power spectrum amplitude. Instead, the allowed 
power spectrum amplitude is simply the minimal range probed by our simulations.

The CV part of our method involves reconstructing the optimal penalty, and thus the strength of the shape prior justified by the data. 
CV is essentially a method to reconstruct the most favoured prior correlation between knots; since the prior is reconstructed 
from the data, prior-driven constraints would not necessarily be a problem. However, here we are finding that no 
particular prior is favoured over any other. Thus, the width of the envelopes in Fig. \ref{fig:current} are actually arbitrary 
and should not be used to draw conclusions about the amplitude of primordial fluctuations at \Lya scales.

We performed a number of checks to determine the cause of our failure to find an optimal penalty. 
Changing our methodology for splitting the data into CV bins did not affect the results. 
A flux power spectrum simulated in the same way as our BOSS data, and using the same parameters, but
with error bars of the same magnitude as the current data showed no preference for a particular penalty, despite, as we shall see, 
there being a well-defined optimal penalty for BOSS simulations. 
Fixing the thermal history parameters $\gamma$ and $T_0$ to fiducial values was also sufficient to allow us to reconstruct a penalty. 
Therefore statistical error and systematic uncertainty in the thermal history are the significant factors preventing us from 
robustly reconstructing a minimally parametric power spectrum shape from current data.

Constraints on the thermal history parameters are as follows. For the low penalty we found $0.8 < \gamma < 1.7$ at 
$1\sigma$ (recall that this upper limit is imposed as a physical prior), while for the high penalty $0.2 < \gamma < 1.7$.
The corresponding constraint from \cite{Viel:2009} is $\gamma = 0.63 \pm 0.5$. There is a noticeable decrease in the best-fitting
value of $\gamma$ with an increased penalty (i.e. a stronger shape prior). 
We find it intriguing that we prefer an inverted temperature-density relation with $\gamma < 1.0$ only for a high penalty, 
but the constraints are
so weak that we cannot draw any solid conclusions from them. 

Constraints on the other parameters at $1\sigma$ were similar for both penalties. Those for the low penalty were: 
$50000 > T_0 > 35000$ K, $\tau_\mathrm{eff} = 0.33 \pm 0.03$, a slope of $\tau^S_\mathrm{eff} = 3.3 \pm 0.3$, $h = 0.7 \pm 0.15$ and $\Omega_M = 0.25 \pm 0.04$.
Finally, constraints on the noise parameters largely reproduce the priors (listed in Sec \ref{sec:mcmcparam}).
Our results mirror those of \cite{Viel:2009}; we have therefore verified that those results are not biased by a shape prior 
on the power spectrum. The constraints of this work and \cite{Viel:2005} on the IGM temperature, $T_0$, prefer a larger
central value than that obtained by \cite{McDonald:2004pk}. However, \cite{McDonald:2004pk} imposed a prior of
$T_0 = 20000 \pm 2000 K$, derived from analysis of the flux probability distribution function of high-resolution quasar spectra, so a direct 
comparison is not possible.
For further discussion of this intriguing result, we refer the interested reader to Section 5 of \cite{Viel:2009}. 

\subsection{Simulated constraints from BOSS}
\label{sec:bossresults}

\begin{figure*}
\centering
\includegraphics[width=0.7\textwidth]{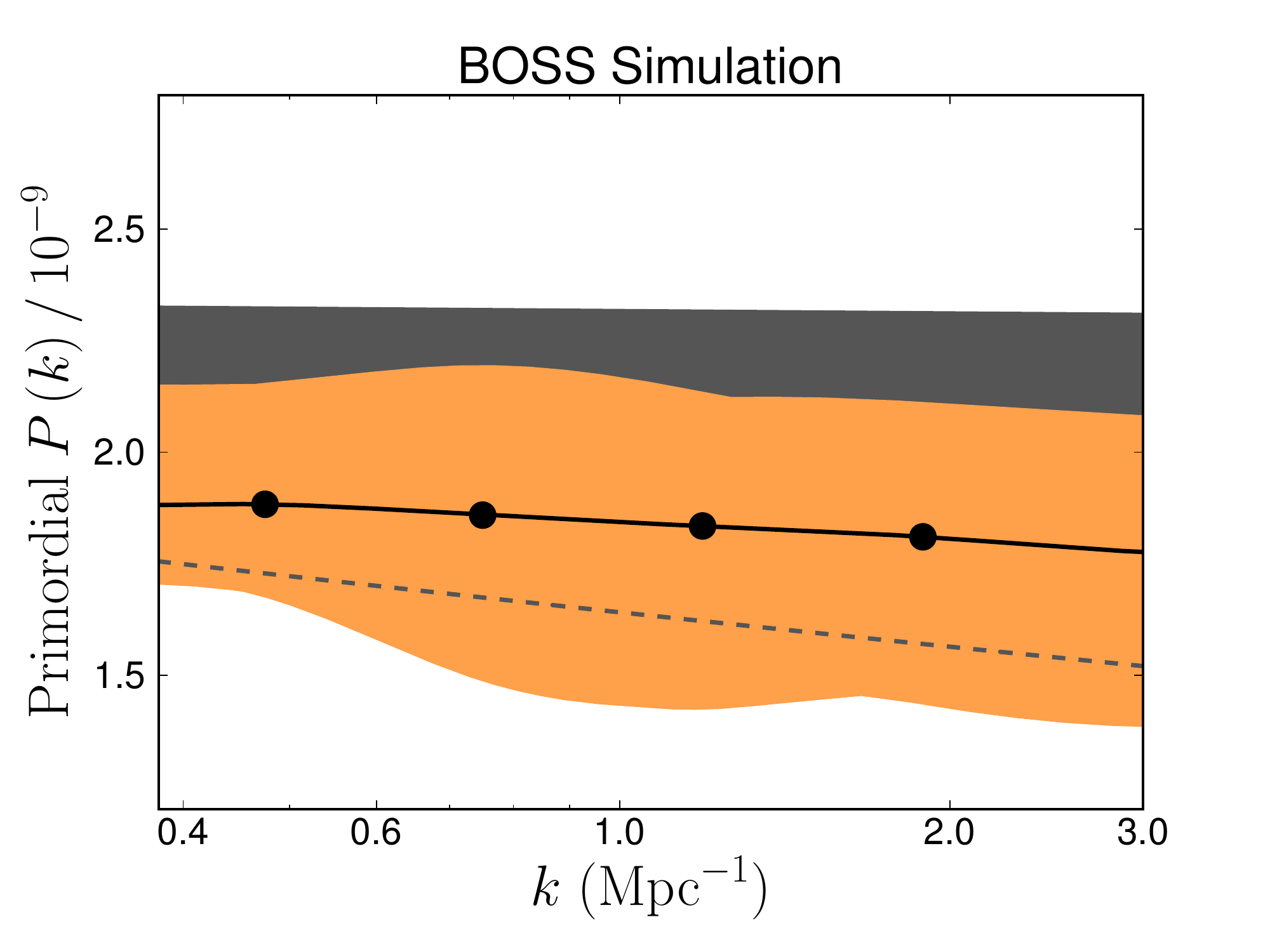} 
\caption{Constraints on the power spectrum for simulated BOSS data. 
Black circles show the positions of the knots, normalized to match 
the input power spectra (black line). 
The orange region shows the top $68\%$ of likelihoods from BOSS-quality \Lya data.
The grey region shows an extrapolation of the $1\sigma$ results from WMAP data to these scales, and the grey dashed line
shows its lower extent.
}\label{fig:boss}
\end{figure*}

\begin{figure*}
\centering
\includegraphics[width=0.9\textwidth]{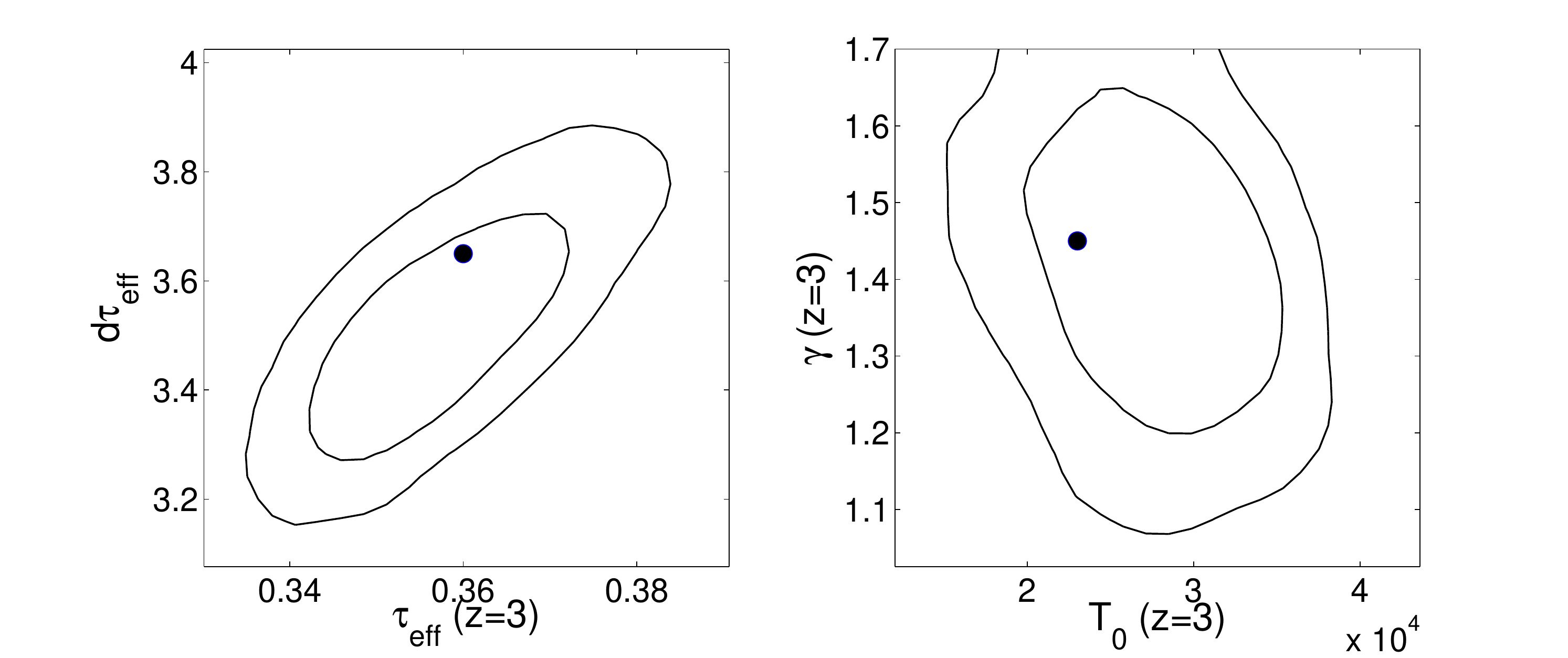} 
\caption{Joint 2D posterior constraints on the thermal history using forecast BOSS data. Input parameters are marked by black dots.
Contours are drawn at $68$ and $95$ percent CL.
See Section \protect\ref{sec:simulations} for definitions of the thermal parameters.}\label{fig:bosstherm}
\end{figure*}

Unlike current data, our simulated BOSS data shows a well-defined maximum in the CV score. 
In Fig. \ref{fig:boss} we show the constraints using this optimal penalty, together with our 
input power spectrum. The input data is reconstructed very well, within an envelope of roughly $0.4 \times 10^{-9}$; 
a precision comparable to that of a CV reconstruction from WMAP data \citep{Verde:2008}. 
Even though our simulated power spectrum is nearly scale-invariant, we do not recover a very high optimal penalty. This is a feature 
of our approach; unless the data is noiseless, not all oscillations in the power spectrum will be ruled out, and the optimal penalty 
is one which allows for them while being consistent with experimental noise.

Our method was designed to extract $P(k)$, and so the penalty may not be entirely optimal for the derivative. 
Even given this, our constraints of $0.7 < n_\mathrm{s} < 1.2$ are still comparatively weak.
However, even this constraint could be useful to test for potential systematics, or in combination with other data sets.
One other important data set will be the power spectrum of the cross-correlation of the 
flux (\citealt{Viel:2002,McDonald:2007, Slosar:2009}), which BOSS is expected to measure for the first time.
Estimating the power of combined constraints is beyond the scope of this paper, but it could be considerable.

Fig. \ref{fig:bosstherm} shows the thermal parameters as reconstructed from BOSS data. We have correctly reconstructed our input, as marked by
the black dots. The reconstructed $h$ and $\Omega_M$ were also consistent with their input values; $\Omega_M = 0.27 \pm 0.02$ (input: $0.267$), 
$h = 0.74 \pm 0.05$ (input: $0.72$). 

Marginalized constraints on the thermal and noise parameters are almost a factor of two better for BOSS than for 
current data. We have assessed the impact that further information about the thermal history of the 
IGM would have on cosmological constraints, imposing priors corresponding to present and reasonable near-future measurements:
\begin{align}
\tau_\mathrm{eff} &= 0.36 \pm 0.11\, ,    &\tau^S_\mathrm{eff} = 3.65 \pm 0.25\, , \nonumber \\
T_0 &= 23000 \pm 3000 K \, ,      &\gamma = 1.45 \pm 0.2\,. \nonumber
\end{align}
Constraints on the mean optical depth are from \cite{Kim:2007}. For the temperature of the IGM, we follow \cite{Becker:2010} 
and assume a future IGM study has determined $\gamma$ to the required precision. 

The effect of the primordial power spectrum evolves with redshift in a different way to $T_0$ and $\gamma$. Hence,
sufficiently accurate data
can break degeneracies between them. For $\tau_\mathrm{eff}$, the constraints from BOSS are already much tighter 
than our prior from \cite{Kim:2007}, so this prior provides no additional information. 
Overall, therefore, the extra information provided by our thermal priors has no significant effect on our reconstruction of 
the primordial power spectrum.

\section{Discussion}
\label{sec:discussion}

\begin{figure*}
\centering
\includegraphics[width=0.7\textwidth]{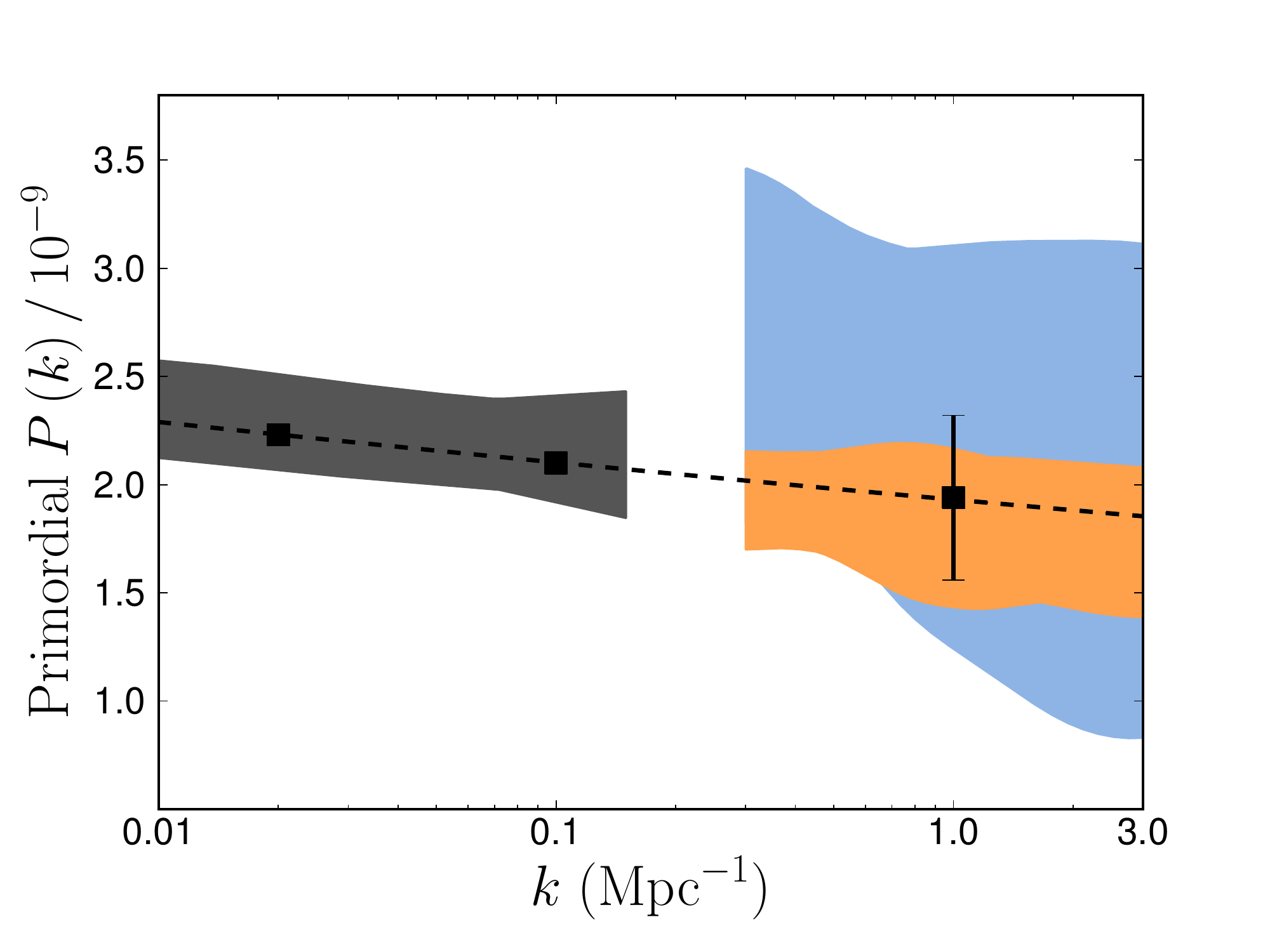} 
\caption{Comparison of our constraints. Blue is from current data; orange is our BOSS forecast.
The grey region shows part of a reconstruction using both the CMB data and galaxy clustering measured 
by SDSS (\protect\citealt{Peiris:2009}). 
The black squares show two knots used in the earlier reconstruction, 
while the black error bar shows $1\sigma$ constraints on power spectrum amplitude from parameter estimation (\protect\citealt{Viel:2009}).
The dashed line shows the extrapolated WMAP best-fitting power spectrum.
}\label{fig:all}
\end{figure*}

In this work, we have performed a minimally parametric reconstruction of the primordial power spectrum, using Lyman-$\alpha$ data. 
This is an extension of \cite{McDonald:2004pk,Viel:2005}, who used Lyman-$\alpha$ data to measure the amplitude 
and slope of the primordial power spectrum on small scales, assuming that it had a power-law shape.
Using a highly prescriptive model to fit data, even if it is physically motivated, can hide systematic effects, which 
may bias the recovered parameters in a manner which is hard to detect unless the bias is extremely large. 
Further, it is vital to go beyond parameter estimation and test the underlying model of the primordial power spectrum. This can in 
principle be achieved with a minimally parametric reconstruction framework coupled with a scheme for avoiding over-fitting the data.

\cite{Peiris:2009}, who attempted such a reconstruction including \Lya data, assumed that 
the power spectrum could be well approximated by an amplitude, a power-law slope and its running across the scales probed by the \Lya forest. In their analysis this assumption was 
justified as the \Lya data was treated as a single point and combined with CMB and galaxy survey data to reconstruct the power spectrum 
over a wide range of scales. 

However, the only likelihood function available up to now contained a power-law assumption 
about the primordial power spectrum shape, making it impossible to treat the \Lya data in a fully 
minimally parametric manner. We remedy this, performing a 
large suite of numerical simulations to construct a new likelihood function. 
The primordial power spectra thus emulated have considerable freedom in their shapes, specified by cubic smoothing splines. 
This provides the first ingredient for a minimally-parametric reconstruction scheme.

The second ingredient, as mentioned above, is to avoid fitting the noise structure of the data with superfluous oscillations.
To this end, our method uses cross-validation (CV) to reconstruct the level of freedom allowed by the data. 
CV is a statistical technique which quantifies the notion that a good fit should be predictive. Schematically, 
it is a method of jackknifing the data as a function of a ``roughness'' penalty. A small penalty thus allows considerable 
oscillatory structure in the power spectrum shape, while a larger penalty specifies a smoother shape. 
This penalty term thus performs the same function as a prior on the smoothness of the power spectrum. 
Jackknifing the data then tests the predictivity of the smoothing prior, choosing as the optimal penalty the one that maximizes 
predictivity. For technical details see Section \ref{sec:pkrecon}.

For the \Lya current data from SDSS (\citealt{McDonald:2004data}), CV yields no significant preference for any 
particular penalty. In the context of CV, this indicates that no penalty is more predictive or favoured over any other; 
in other words, the data are not sufficiently powerful to accurately reconstruct the strength of the shape prior. 

The minimally parametric method thus provides no evidence for features in the power spectrum in the current data, and our results are 
fully consistent with a scale-invariant power spectrum. The best-fitting amplitude 
of the power spectrum is, as in previous work, slightly higher than that extrapolated from WMAP (\citealt{WMAP7}). 
However, because the data do not contain sufficient statistical power to reconstruct the power spectrum shape, our error 
bars are extremely large. An analysis that uses different statistical 
techniques, such as Bayesian evidence (\citealt{Jeffreys:1961}), could provide further insight, but is beyond the scope of this paper.

In the not so distant future, the first data from a new Lyman-$\alpha$ survey, BOSS (\citealt{Schlegel:2009}),
will be made available. We simulate a flux power spectrum and covariance matrix for BOSS, with an 80 fold 
increase in statistical power over the current data. 
In this case we successfully reconstruct the power spectrum, using CV to find an optimal penalty. 
The parameters we extract using CV are completely consistent with the inputs to the simulation, and the resulting constraints are comparable 
to those achieved by performing CV reconstruction using WMAP data (\citealt{Verde:2008}). We verify that statistical error is the 
factor preventing us from finding an optimal penalty for current data by simulating a power spectrum identical to BOSS, but 
with wider error bars, again failing to find an optimal penalty. 

Finally, we show that adding plausible future data on the smallscale thermodynamics of the IGM to BOSS does not significantly 
improve constraints on the primordial power spectrum. The simulated BOSS data are sufficiently powerful on their own to break degeneracies 
between the IGM and cosmological parameters, and are limited by statistical error rather than systematic uncertainty. 

We have not considered the impact of the information BOSS is expected to provide on the transverse flux power spectrum. 
This will probe larger scales than our current work, offering a longer baseline and thus better sensitivity to the overall 
shape of the power spectrum. However, applying the present technique to the improved data set would require simulations 
probing much larger scales, hence greatly increasing the numerical requirements.

Fig. \ref{fig:all} shows the constraints from BOSS in comparison to those \cite{Peiris:2009} obtained by reconstructing the power 
spectrum using the CMB and the matter power spectrum from SDSS. By combining BOSS data with other probes
(\citealt{Seljak:2005, Seljak:2006}), such as galaxy clustering, the CMB, 
and the transverse flux power spectrum, we will be able to accurately reconstruct the 
shape of the power spectrum on scales of $k=0.001 - 3 \,\Mpc^{-1}$, probing ten $e$-folds 
of inflation. 


\section*{Acknowledgments}

This work was performed using the Darwin Supercomputer of the University of Cambridge High 
Performance Computing Service (http://www.hpc.cam.ac.uk/), provided by Dell Inc. 
using Strategic Research Infrastructure Funding from the Higher Education 
Funding Council for England.
We thank Volker Springel for writing and making public the {\small GADGET}-2 code, and 
for giving us permission to use his initial conditions code N-GenICs.  
SB would like to thank Martin Haehnelt, Antony Lewis, Debora Sijacki, and Christian Wagner
for help and useful discussions. 
SB is supported by STFC. HVP is supported by Marie Curie grant MIRG-CT-2007-203314 from the 
European Commission, the Leverhulme Trust, and by STFC.
MV is partly supported by: ASI/AAE theory grant, INFN-PD51 grant,  PRIN-MIUR
and a PRIN-INAF and an ERC starting grant.
LV acknowledges support of MICINN grant AYA2008-03531 and FP7-IDEAS-Phys.LSS 240117 
and thanks IoA Cambridge for hospitality.


\appendix
\section{Simulated Spectra}\label{ap:absorption}

In this appendix we detail the procedure for extracting a spectrum from a simulation snapshot.
First, we must find the velocity of each particle, including both peculiar velocities and the Hubble flow. 
The effect of peculiar velocities is to increase the flux power by around $10\%$. 

Next, we calculate the optical depth at wavelength $\lambda$, as defined by the line integral
\begin{equation}
        \tau_\lambda = \int \sigma_\lambda n_H dl \,,
        \label{eq:opt}
\end{equation}
where $\sigma_\lambda$ is the cross-section for the transition and $n_H$ is the number density of the neutral hydrogen. $\sigma_\lambda$ is given by
the rest cross-section multiplied by a broadening function
\begin{equation}
        a_\lambda = \sigma_\alpha \times \Phi\,.
\end{equation}
We define the oscillator strength, $f_\alpha$, by the ratio between the cross-section of the \Lya transition and the 
cross section of the transition involving a free electron 
\begin{equation}
        f_\alpha = \frac{\sigma_\alpha}{\pi r_0\lambda}\,,
\end{equation}
where $\lambda$ is the rest wavelength of the Lyman-$\alpha$ transition.
The classical radius of the electron, $r_0$ is related to the Thompson 
cross-section, $\sigma_T$ by
\begin{equation}
        r_0 = \sqrt{\frac{3\sigma_T}{8\pi}} \,.
\end{equation}
Hence 
\begin{equation}
        \sigma_\alpha = \sqrt{\pi} \sqrt{\frac{3\sigma_T}{8}} f_\alpha \lambda\,.
        \label{eq:lyacross}
\end{equation}

To compute the broadening function, we neglect natural broadening from the 
intrinsic uncertainty in the energy levels of the hydrogen atom. Natural broadening 
is only important in the densest absorbers (damped \Lya systems), which our 
simulations lack the resolution to adequately resolve. The effect of DLAs is included 
by a correction applied when calculating the likelihood, for which see section \ref{sec:mcmcparam}.

Hence the only form of broadening present is Doppler broadening. In an absorber  of 
temperature $T_H$, mass $m_H$, and velocity $v$, the probability of a particle
having zero velocity relative to an incoming photon is
\begin{align}
        \Phi &= \frac{c}{\sqrt{\pi} b} \exp \left[-\frac{v}{b}^2\right]\,, \\
        \mathrm{where}\; b &= \sqrt{\frac{2 k T_H}{m_H}} \,.\nonumber
        \label{eq:broaden}
\end{align}
Hence, a wavelength bin at position $k$ will suffer absorption from a $HI$ absorber in
a bin $j$ as:
\begin{align}
        \tau_{kj} &= \sigma_\alpha \Phi n_H a \Delta \\
        &= \sigma_\alpha \frac{c}{\sqrt{\pi} b} n_H a \Delta 
        \exp\left[-\left(\frac{v_k -v_j}{b}\right)^2\right]\,.
        \label{eq:optical}
\end{align}
Here $\Delta$ is the bin width, and $a$ is the expansion factor. 

The flux in each bin is then simply $\mathcal{F} = e^{-\tau}$. Each spectra is smoothed with a simple
$3$ point boxcar average, following \cite{Viel:2004}, and the flux power spectrum from the simulation box 
is defined to be the average over a number of simulated spectra. 

\section{Convergence Checks}
\label{ap:converge}

In this appendix, we detail the checks we have performed to ensure that our simulations are properly converged with respect to box size and 
particle number. We have usually been comparing the relative change in the flux power spectrum when changing a parameter, 
making strict convergence not essential.

To check box-size convergence, we compared the flux $P_F(k)$ for our fiducial simulation with a large box size simulation (``L'') which 
had size ($75$, $500$), and otherwise identical parameters to the fiducial simulation (``F''). This isolates the effect of box size 
by having identical particle resolution to simulation F. To test convergence with respect to particle number, we used a high resolution 
simulation (``H''), with ($60$, $500$). In order to isolate the change due to numerical effects, we did not rescale the mean optical depth
for the plots in this section.

\begin{figure*}
\centering
\includegraphics[width=0.4\textwidth]{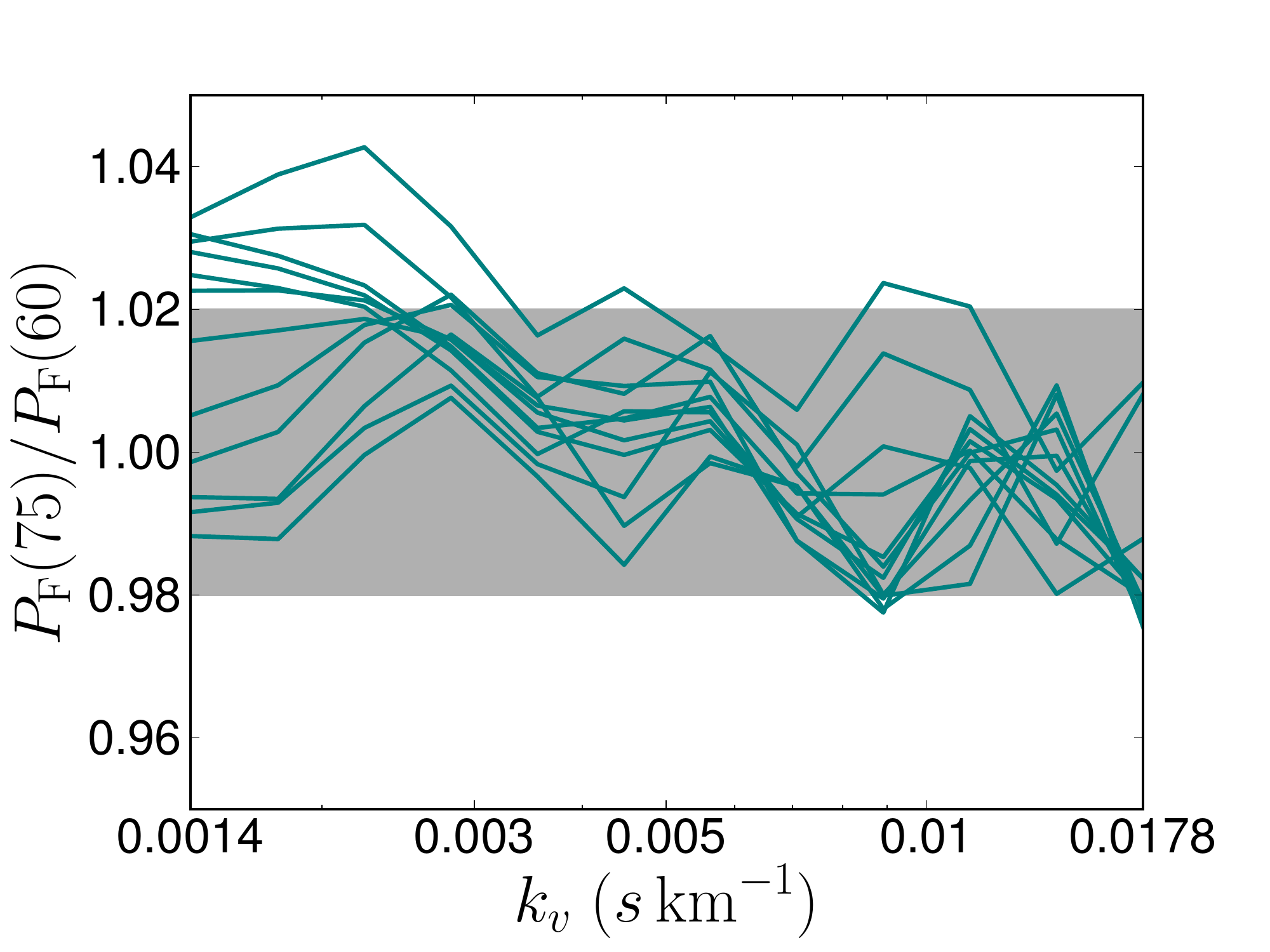} \includegraphics[width=0.4\textwidth]{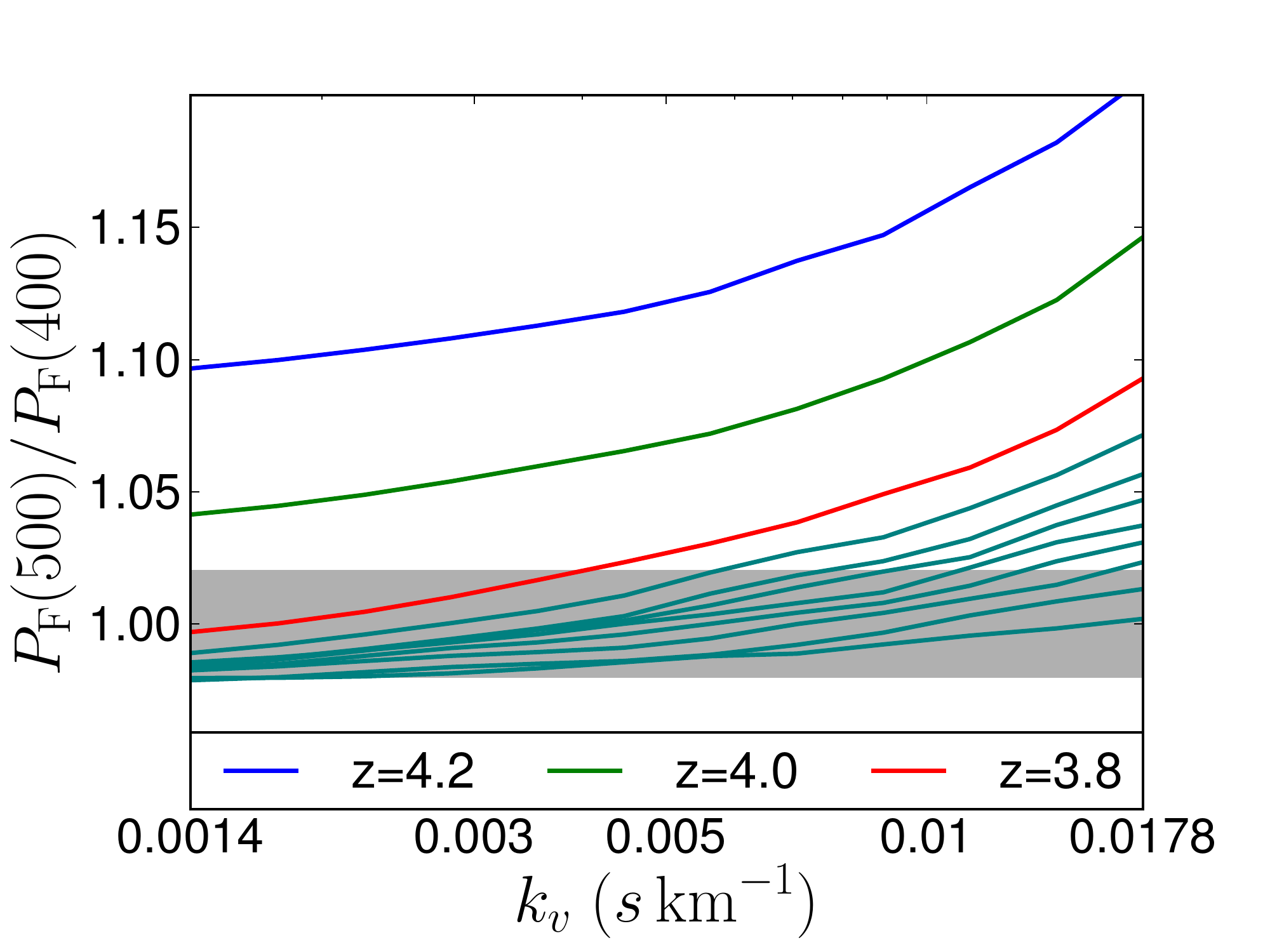}
\caption{(Left) The change in the flux power spectrum due to increasing the box size at fixed particle resolution. Each green line shows the effect on a different redshift bin, from $4.2$ to $2.0$. The effect is generally around $2\%$ (grey box), with a systematic increase on large scales, for which we correct (see text).
(Right) The effect on the flux power spectrum due to increasing the particle number from $2\times 400^3$ to $2\times 500^3$. Each line shows the effect on a different redshift bin, from $4.2$ to $2.0$. The grey box show a variation of $4\%$. Redshift bins with greater variation are labelled. 
}\label{con:boxres}
\end{figure*}

The left hand plot of Fig. \ref{con:boxres} shows the change in the flux power spectrum with increased box size; 
simulation L divided by simulation F.
The flux power spectrum is converged with respect to box-size; however, there is a systematic increase on large scales. 
This is due to sample variance: the specific realisation of cosmic structure we are using is biased slightly low on the 
largest scales probed by the box. The larger box recovers the input power spectrum much better on these scales, 
because it contains far more modes, and hence shows an increase in power. Because we use the same realisation of 
structure for all our simulations, this effect will be constant, and is easily corrected for by altering the best-fitting 
power spectrum. Once this is done, convergence of the flux power spectrum is very good.

The right hand plot of Fig. \ref{con:boxres} shows the change in the flux power spectrum with increased particle resolution. 
The effect is small, except on small scales or at high redshift. Achieving numerical convergence for the \Lya forest at high redshift
is challenging, because most of the signal for the \Lya forest is coming from poorly resolved under-dense regions.
In addition, current data at high redshifts is much more noisy than at low redshifts, 
due to a paucity of quasar spectra. Accordingly, we follow \cite{Viel:2005} and do not use the three highest redshift bins 
at $z=4.2,4.0$ and $3.8$ in our analysis. 

Our initial base for the central power spectrum was the output of simulation H. 
We corrected for box size and resolution following the method proposed by \cite{McDonald:2003}.
To correct for sample variance on the largest scales, we ran two additional simulations with box size $120\,\Mpch$, and 
 $400^3$ and $200^3$ particles. The smaller simulation has identical particle resolution to our fiducial simulations, 
 and should thus be directly comparable to them.
 We then corrected our best-guess power spectrum by the ratio between the two,
\begin{equation}
        C_S(k) = P_F(N_P=400)/P_F(N_P=200)\,.
        \label{eq:boxcorrection}
\end{equation}
The smallest scale bin was excluded from this correction as it was clearly being affected by poor resolution convergence.

Because simulation H is nearly converged, we had to be careful when correcting for resolution; if the error in the correction 
is larger than the correction itself, accuracy is definitely not increased. Therefore, 
we ran two simulations, with box size of $24\,\Mpch$; $T1$ and $T2$. 
$T1$ has the same particle resolution as simulation H, and thus $200^3$ particles.
$T2$ has $400^3$, giving it much increased particle resolution. These simulations 
do not resolve the largest scales probed by the \Lya forest at all, but Fig. \ref{con:boxres} 
shows that these scales are not affected by poor resolution convergence. 
Simulation H is corrected by 
\begin{equation}
        C_R(k) = P_F^{T2}/P_F^{T1}\,.
        \label{eq:rescorrection}
\end{equation}
To avoid our correction itself being biased by a small box size, we ignore those bins
on scales greater than a quarter of the box-size of $24\Mpch$. We also ignore any correction
for redshift bins where the correction for the smallest scale bin is less than the uncertainty
in the simulations, which we take to be $1\%$, on the grounds that these are 
already fully converged. We are left with a slight increase in power on the smallest scales.

\bibliographystyle{mn2e_arxiv}
\bibliography{paper}


\label{lastpage}
\end{document}